\documentclass{aip-cp}

\usepackage[numbers]{natbib}
\usepackage{rotating}
\usepackage{graphicx}

\def\Dtilde{\tilde{D}}
\def\Etilde{\tilde{E}}
\def\Ptilde{\tilde{P}}
\def\ftilde{\tilde{f}}

\def\aprime{a^\prime}
\def\abis{a^{\prime\prime}}

\def\rperp{{\bf r}_\perp}
\def\kperp{k_\perp}
\def\pperp{p_\perp}

\def\kperpone{k_{1 \perp}}
\def\kperptwo{k_{2 \perp}}
\def\kperponeb{\mathbf{k}_{1 \perp}}
\def\kperptwob{\mathbf{k}_{2 \perp}}
\def\kperpthree{k_{3 \perp}}
\def\kperpthreeb{\mathbf{k}_{3 \perp}}
\begin{document}

\title{Unintegrated Double Parton Distributions - a Summary}

\author[aff1,aff2]{Krzysztof Golec-Biernat}
\eaddress{golec@ifj.edu.pl}
\author[aff3]{Anna Sta\'sto}
\eaddress{ams52@psu.edu}

\affil[aff1]{Institute of Nuclear Physics, Polish Academy of Sciences, 31-342 Cracow, Poland}
\affil[aff2]{Faculty of Mathematics and Natural Sciences, University of Rzesz\'ow,  35-959 Rzesz\'ow, Poland}
\affil[aff3]{Department of Physics, The Pennsylvania State University, University Park, PA 16802, United States}

\maketitle

\begin{abstract}
We present main elements of the construction of unintegrated double parton distribution functions which depend on   transverse momenta of partons. We follow the method proposed by Kimber, Martin and Ryskin for a construction of unintegrated single parton distributions from the standard parton distribution functions. 
\end{abstract}

\section{INTRODUCTION}

Double parton scattering belongs to a class of multi-parton interactions in hadronic scattering in which two systems with hard scales are produced in one event due to two independent
parton-parton interactions, see \cite{Diehl:2011yj} for a theoretical introduction. Such events were observed by  both the Fermilab \cite{Abe:1997bp,Abe:1997xk,Abazov:2009gc} 
and CERN experiments \cite{Aad:2013bjm,Chatrchyan:2013xxa,Aad:2014rua}. The standard QCD description of such processes is based on double parton distribution functions (DPDFs) and collinear factorization of cross sections, which is a generalization of the single parton scattering description with the well known single  parton distribution functions (PDFs). 
We present a first attempt to a further generalization in which the DPDFs start to depend on transverse momenta of active partons, thus may be used in $k_\perp$-factorized
cross sections with off-shell partons initiating hard scatterings.  We follow the method proposed in References \cite{Kimber:1999xc,Kimber:2001sc} for the construction of  unintegrated
PDFs by unfolding the last step in the QCD evolution of PDFs. This is why we call our distributions unintegrated DPDFs (UDPDFs). The full documentation of our construction can be found in Reference \cite{Golec-Biernat:2016vbt}.

\section{DOUBLE PARTON DISTRIBUTIONS}

\begin{figure}[t]
\centerline{
{\includegraphics[width=0.22\textwidth]{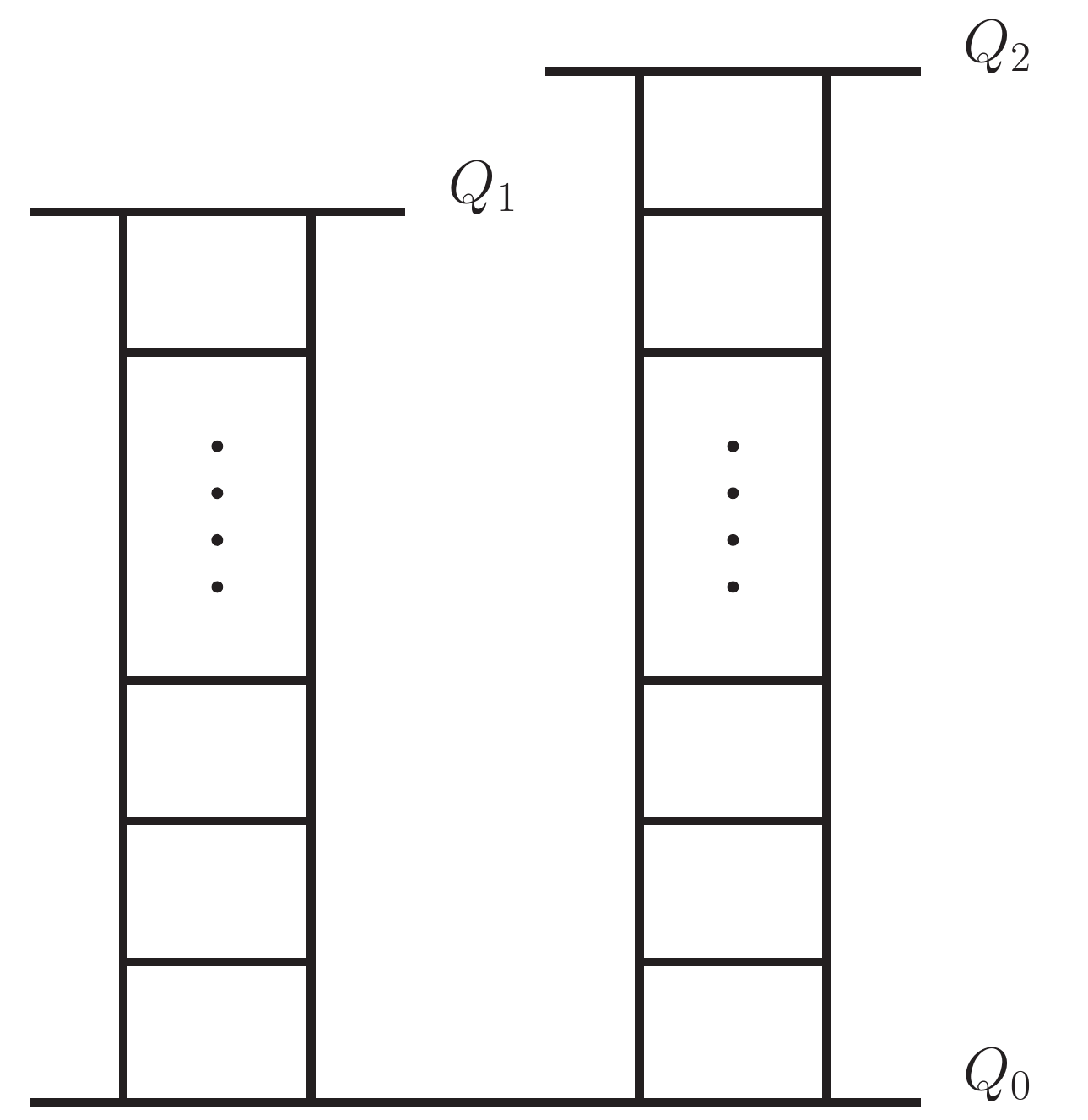}}\hspace*{1.7cm}
{\includegraphics[width=0.22\textwidth]{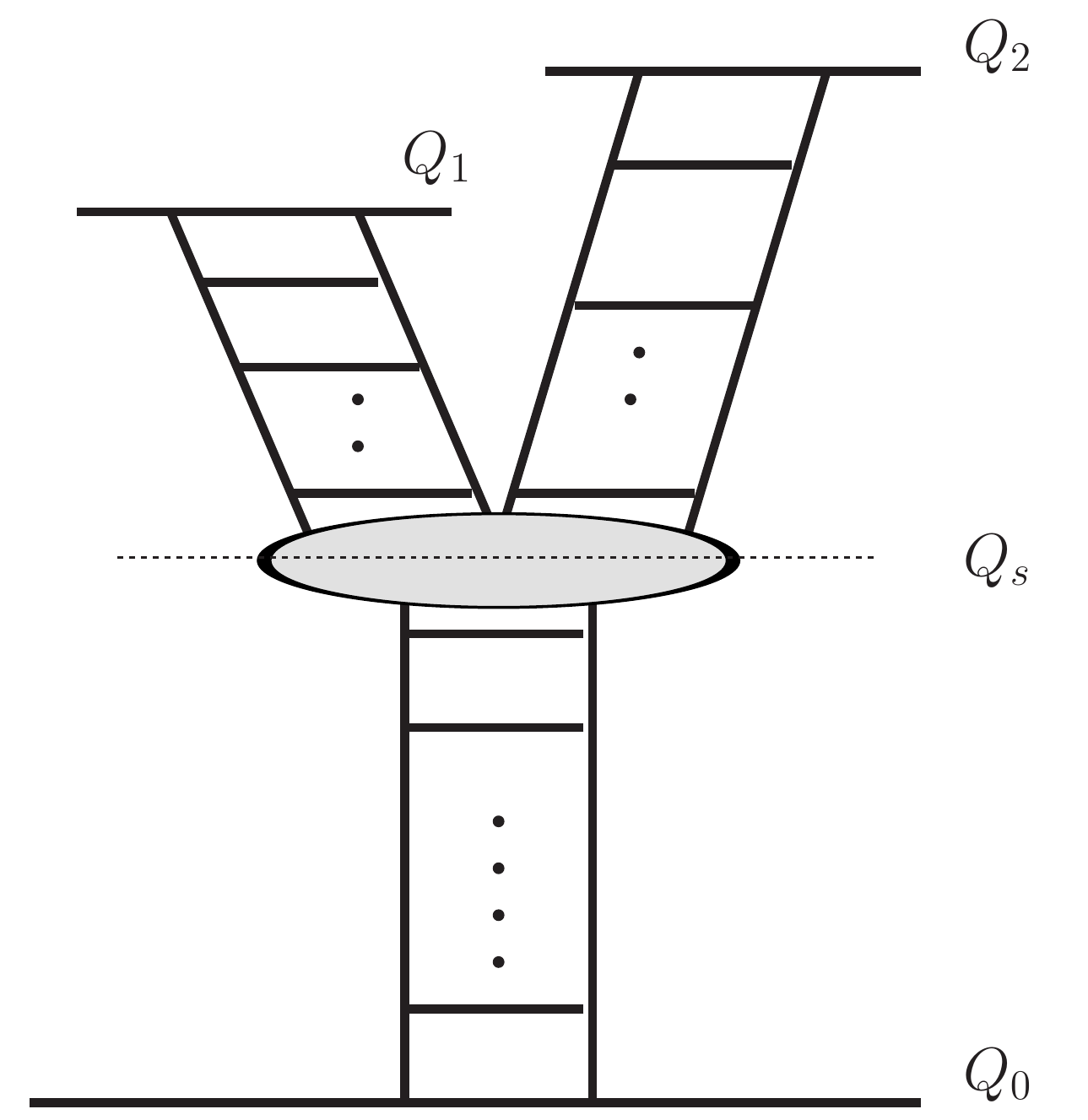}}  
}
\caption{Schematic illustration of   two contributions to the evolution of DPDFs, see Equation~\ref{eq:2}.}
\label{fig:1}
\end{figure}

The DPDFs, $D_{a_1a_2}(x_1,x_2,Q_1,Q_2)$, depend on positive, longitudinal momentum fractions, $x_{1,2}$, of two partons of flavors (including gluon) $a_{1,2}$, and also on two
hard scales $Q_{1,2}$. To save on space we switch to their double Mellin moments
\begin{equation}
\label{eq:1}
\Dtilde_{a_1a_2}(n_1,n_2,Q_1,Q_2)=\int_0^1dx_1\int_0^1dx_2\,x_1^{n_1}x_2^{n_2}\,\theta(1-x_1-x_2)\,D_{a_1a_2}(x_1,x_2,Q_1,Q_2)\,.
\end{equation}
In the leading logarithmic approximation, the DPDFs evolve with hards scales according the equation \cite{Snigirev:2003cq,Ceccopieri:2010kg}
\begin{eqnarray}\nonumber
\Dtilde_{a_1a_2}(n_1,n_2,Q_1,Q_2) &=& \sum_{\aprime,\abis}\bigg\{
\Etilde_{a_1\aprime}(n_1,Q_1,Q_0)\,\Etilde_{a_2\abis}(n_2,Q_2,Q_0)\,
\Dtilde_{\aprime\abis}(n_1,n_2,Q_0,Q_0) 
\\
&+& \int^{Q_{\min}^2}_{Q_0^2}\frac{dQ^2_s}{Q^2_s}\,
\Etilde_{a_1\aprime}(n_1,Q_1,Q_s)\,\Etilde_{a_2\abis}(n_2,Q_2,Q_s)\,
\Dtilde^{(sp)}_{\aprime\abis}(n_1,n_2,Q_s)\bigg\} \; ,
\label{eq:2}
\end{eqnarray}
where $Q^2_{min}={\rm min}\{Q^2_1,Q^2_2\}$, $\Dtilde_{\aprime\abis}(n_1,n_2,Q_0,Q_0)$ is an initial condition and
\begin{equation}
\label{eq:3}
\Dtilde^{(sp)}_{\aprime\abis}(n_1,n_2,Q_s)=\frac{\alpha_s(Q_s)}{2\pi}\sum_a \Dtilde_a(n_1+n_2,Q_s)
\int_0^1dz\,z^{n_1}(1-z)^{n_2}P_{a\to\aprime\abis}(z)\,.
\end{equation}
In Equation~\ref{eq:3} 
$P_{a\to\aprime\abis}(z)\equiv P_{\aprime a}(z)$ is the leading order Altarelli-Parisi splitting function
and  $\Dtilde_a(n_1+n_2,Q_s)$ is the single PDF.
The structure of Equation~\ref{eq:2} is illustrated in Figure~\ref{fig:1}, where the first (homogeneous) term in this equation corresponds to  the left picture while the second (splitting) term corresponds to the right picture. $Q_s$ in the splitting term is the scale where the parton splitting occurs. Thus, the two partons originate either from a nucleon or from the perturbative parton splitting.
The ladder diagrams represent the evolution functions, $\Etilde_{ab}(n,Q,Q_0)$, which
are the DGLAP parton in parton distribution  functions, obeying the following 
integral equation\footnote{This equation can be readily recast into the differential form of the DGLAP equation with the initial condition $\Etilde_{ab}(n,Q_0,Q_0)=\delta_{ab}.$}
\begin{equation}
\label{eq:4}
\Etilde_{ab}(n,Q,Q_0) = T_a(Q,Q_0)\,\delta_{ab}+
\int_{Q_0^2}^{Q^2}  \frac{d\kperp^2}{\kperp^2}\,T_a(Q,\kperp)
\sum_{\aprime}\Ptilde_{a\aprime}(n,\kperp)\,\Etilde_{\aprime b}(n,\kperp,Q_0)\,,
\end{equation}
where $\Ptilde_{a\aprime}(n,\kperp)$ is the Mellin transformed splitting function with $\kperp$-dependence in the strong coupling constant, and 
\begin{equation}
\label{eq:5}
T_a(Q,\kperp)=\exp\bigg\{-\int_{\kperp^2}^{Q^2}\frac{d\pperp^{2}}{\pperp^{2}}\sum_{\aprime}\int_0^1dz \,z\, P_{\aprime a}(z,\kperp)\bigg\}
\end{equation}
is the Sudakov formfactor which sums virtual corrections. 
 
\section{UNINTEGRATED DOUBLE PARTON DISTRIBUTIONS}

We have already introduced all the elements to unfold the transverse momentum dependence from Equation \ref{eq:2}. In the DGLAP scheme,
the $\kperp$ in Equation \ref{eq:4} is the transverse momentum of the $t$-chanel (exchanged) parton. Thus, after substituting (\ref{eq:4}) into (\ref{eq:2}),
we unfold the $\kperp$-dependence from the DPDFs as integrands of the $\kperp$ integrations. 

\begin{figure}[t]
\includegraphics[width=0.30\textwidth]{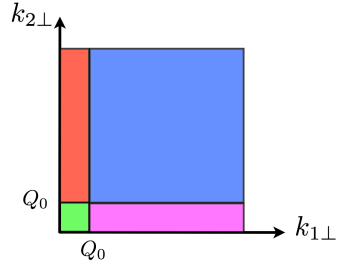} 
\caption{Regions of transverse momenta for the homogeneous part of the UDPDFs, defined through Equation~\ref{eq:6}.}
\label{fig:2}
\end{figure}

For the first, homogeneous term in Equation~\ref{eq:2}, we find after
such a  substitution
\begin{eqnarray}\nonumber
\label{eq:6}
&&\Dtilde^{(h)}_{a_1a_2}(n_1,n_2,Q_1,Q_2) = T_{a_1}(Q_1,Q_0)\,T_{a_2}(Q_2,Q_0)\,\Dtilde_{a_1a_2}(n_1,n_2,Q_0,Q_0)
\\\nonumber
&+& \int_{Q_0^2}^{Q_2^2}  \frac{d\kperptwo^2}{\kperptwo^2}
\bigg\{T_{a_1}(Q_1,Q_0)\,T_{a_2}(Q_2,\kperptwo)\sum_{b}\Ptilde_{a_2b}(n_2,\kperptwo)\,\Dtilde^{(h)}_{a_1b}(n_1,n_2,Q_0,\kperptwo)\bigg\}
\\\nonumber
&+& \int_{Q_0^2}^{Q_1^2}  \frac{d\kperpone^2}{\kperpone^2}
\bigg\{T_{a_1}(Q_1,\kperpone)\,T_{a_2}(Q_2,Q_0)\sum_{b}\Ptilde_{a_1b}(n_1,\kperpone)\,\Dtilde^{(h)}_{b a_2}(n_1,n_2,\kperpone,Q_0)\bigg\}
\\
&+&
\int_{Q_0^2}^{Q_1^2}  \frac{d\kperpone^2}{\kperpone^2}
\int_{Q_0^2}^{Q_2^2}  \frac{d\kperptwo^2}{\kperptwo^2}
\bigg\{T_{a_1}(Q_1,\kperpone)\,T_{a_2}(Q_2,\kperptwo)
\sum_{b,c}\Ptilde_{a_1b}(n_1,\kperpone)\,\Ptilde_{a_2c}(n_2,\kperptwo)\,
\Dtilde^{(h)}_{bc}(n_1,n_2,\kperpone,\kperptwo)\bigg\},~~~~~
\end{eqnarray}
where the functions $\Dtilde^{(h)}$ on the right hand side are the homogeneous DPDFs evolved from initial conditions like on the left picture 
in Figure~\ref{fig:1}, e.g
\begin{equation}
\label{eq:7}
\Dtilde^{(h)}_{bc}(n_1,n_2,\kperpone,\kperptwo) = \sum_{\aprime,\abis}\Etilde_{b \aprime}(n_1,\kperpone,Q_0)\,
\Etilde_{c \abis}(n_2,\kperptwo,Q_0)\,\Dtilde_{\aprime\abis}(n_1,n_2,Q_0,Q_0)\,.
\end{equation}
The integrands in the curly brackets in Equation \ref{eq:6} are the unintegrated DPDFs, $\ftilde^{(h)}_{a_1a_2}$, which are defined in three regions of the transverse momentum 
plane, $(\kperpone, \kperptwo)$, shown in Figure~\ref{eq:2} as the red, pink and blue rectangles. The first two expressions (from the top)  in Equation \ref{eq:6}
have one of the two transverse momenta equal to the initial scale  $Q_0$. It means
that this transverse momentum is integrated up to $Q_0$ and is not present
among the arguments of the defined function. The effect of such an integration is hidden in the integrated DPDFs, $\Dtilde^{(h)}$, 
taken at  $Q_0$ for one of the two scales. Such UDPDFs correspond to the red and pink regions in Figure~\ref{eq:2}.  The UDPDF in the blue region is defined by the third
term in the curly brackets in Equation \ref{eq:6}, which after transforming back to the $x$-space reads
\begin{eqnarray}\nonumber
\label{eq:8}
f_{a_1a_2}^{(h)}(x_1,x_2,\kperpone,\kperptwo, Q_1,Q_2) &=&
T_{a_1}(Q_1,\kperpone)\,T_{a_2}(Q_2,\kperptwo)
\\
&\times&
\sum_{b,c}
\int_{\frac{x_1}{1-x_2}}^{1-\Delta_1}\frac{dz_1}{z_1}
\int_{\frac{x_2}{1-x_1/z_1}}^{1-\Delta_2}\frac{dz_2}{z_2}
P_{a_1b}(z_1,\kperpone)\,
P_{a_2c}(z_2,\kperptwo)\,
D^{(h)}_{bc}\Big(\frac{x_1}{z_1},\frac{x_2}{z_2},\kperpone,\kperptwo\Big)\,.~~~~
\end{eqnarray}
The upper limits in the integrals above are shifted from $1$ by
$\Delta_i=k_{i\perp}/Q_i$ for $i=1,2$
to regularize the divergence of the flavor diagonal splitting functions at $z=1$. The same procedure is applied to the Sudakov formfactor (\ref{eq:5}).
A closer inspection of Equation~\ref{eq:8} shows that the longitudinal and transverse momenta  are correlated by the relation
\begin{equation}
\label{eq:9}
\frac{x_1}{1-\Delta_1}+\frac{x_2}{1-\Delta_2}\le 1\,,
\end{equation}
which is a stronger condition than that for the DPDFs, $x_1+x_2\le 1$. Finally, the green region in Figure~\ref{fig:2} corresponds to the first term
in Equation~\ref{eq:6} in which both transverse momenta are integrated up to the scale $Q_0$. Thus, there are no UDPDFs in this region. 

In principle, all the regions of  transverse momenta need to be included for any  configuration of the external hard scales $Q_1$ and $Q_2$. It is clear though, that some regions will be subdominant depending on the scales, due to the suppression originating from the Sudakov formfactors. For example,
the first term in Equation~\ref{eq:6} is going to be very small whenever any of the scales is much larger than $Q_0$.

\section{DISCUSSION OF THE SPLITTING CONTRIBUTION}

\begin{figure}[t]
\centerline{
{\includegraphics[width=0.30\textwidth]{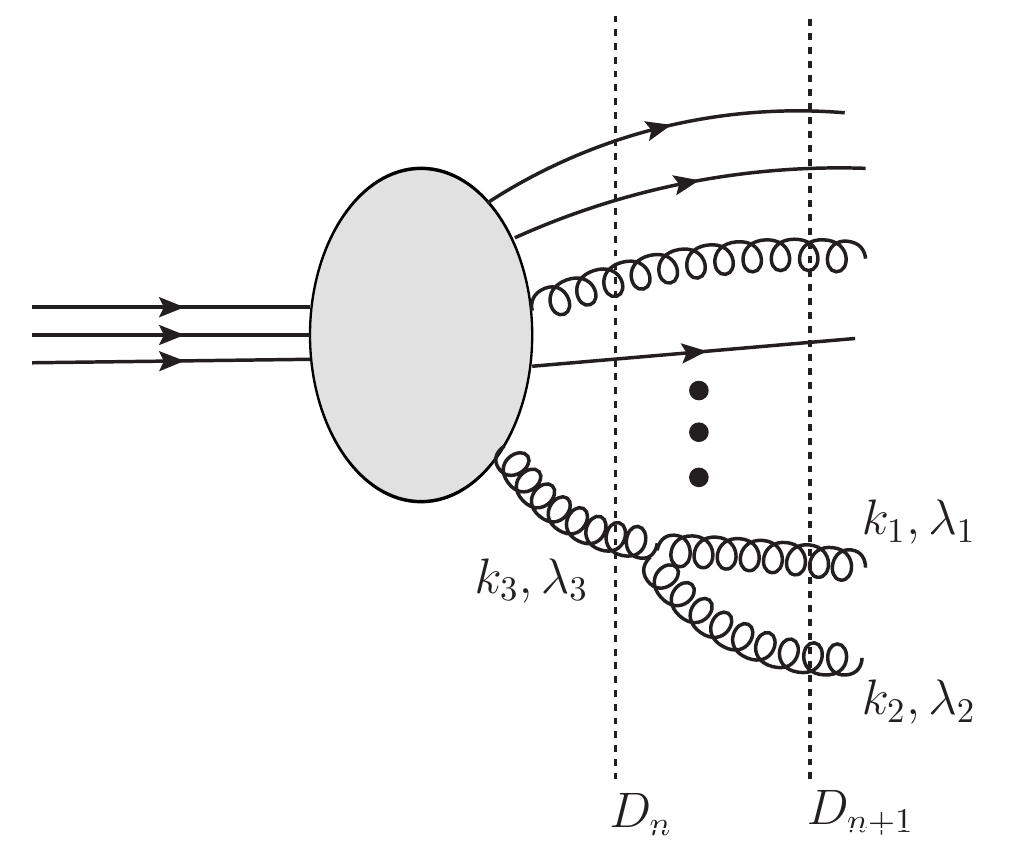}} 
}
\caption{Splitting contribution to the proton wave function in the light-front framework.}
\label{fig:3}
\end{figure}

The discussion of the splitting contribution is more involved since in principle there are two potential sources of transverse momenta dependence in this case,  from the splitting vertex itself  and  from the evolution above the splitting vertex. 
In the latter case, we can apply the method from the previous section to the second term in Equation~\ref{eq:2}. 
For example, in the blue region in Figure~\ref{eq:2}, i.e. for $\kperpone,\kperptwo\ge Q_0$, we find the following unintegrated DPDFs in the $x$-space from the splitting mechanism \cite{Golec-Biernat:2016vbt}
\begin{eqnarray}\nonumber
f^{(sp)}_{a_1a_2}(x_1,x_2,\kperpone,\kperptwo,Q_1,Q_2) &=&
T_{a_1}(Q_1,\kperpone)\,T_{a_2}(Q_2,\kperptwo)
\int_{\frac{x_1}{1-x_2}}^{1-\Delta_1}\frac{dz_1}{z_1}
\int_{\frac{x_2}{1-x_1/z_1}}^{1-\Delta_2}\frac{dz_2}{z_2}\,\sum_{b,c}
P_{a_1b}(z_1,\kperpone)
P_{a_2c}(z_2,\kperptwo)
\\
&\times&
\int_{Q_0^2}^{Q_{\min}^2}\frac{dQ_s^2}{Q_s^2}\,
\theta(k_{1\perp}^2-Q_s^2)\,\theta(k_{2\perp}^2-Q_s^2)\,
{\cal D}^{(sp)}_{bc}\Big(\frac{x_1}{z_1},\frac{x_2}{z_2},\kperpone,\kperptwo,Q_s\Big)\,,
\label{eq:10}
\end{eqnarray}
where the integrated distribution ${\cal D}^{(sp)}_{bc}$ on the right hand side is obtained from the two ladder evolution, i.e. in the Mellin moment space
\begin{equation}
\label{eq:11}
\tilde{\cal D}^{(sp)}_{bc}(n_1,n_2,\kperpone,\kperptwo,Q_s)
=\sum_{\aprime,\abis}\Etilde_{b\aprime}(n_1,\kperpone,Q_s)\, \Etilde_{c\abis}(n_2,\kperptwo,Q_s)\,
\Dtilde^{(sp)}_{\aprime\abis}(n_1,n_2,Q_s)\,.
\end{equation}
By the comparison with Equation~\ref{eq:7} we see that the "initial condition" for this evolution is given by the effective distribution (\ref{eq:3}) which contains
the single PDFs taken at the splitting scale $Q_s$. The analysis of the two remaining contributions, in which one of the two momenta is integrated out, reveals that
they cannot be treated on the same footing as those in the homogeneous case. The reason is the lack of a clear cut division between
the perturbative and non-perturbative regions since the integration over the transverse momentum extends up to $Q_s$ which can be much bigger than $Q_0$ \cite{Golec-Biernat:2016vbt}. 
Thus only formula (\ref{eq:10}), valid for  $\kperpone,\kperptwo\ge Q_0$, is acceptable in the discussed case.

For the discussion of the  transverse momentum dependence coming directly from the 
perturbative splitting of a single parent parton into two daughter partons,  
we utilized the methods of the light-front perturbation theory. Applying these methods to the diagram shown in Figure~\ref{fig:3} we were able to find the result  (\ref{eq:3})
for  the splitting term in the evolution equations (\ref{eq:2}) for the integrated DPDFs.  Going deeper into the transverse momentum dependence of the expressions leading to 
this results, we found in the strong ordering approximation of transverse momenta, $\kperp\simeq \kperpone \simeq \kperptwo  \gg \kperpthree$, the following unintegrated 
DPDFs form the splitting vertex
\begin{equation}
\label{eq:12}
f_{a_1 a_2}(x_1,x_2,\kperponeb,\kperptwob) =  \frac{\alpha_s}{2\pi}  \frac{1}{x_1+x_2}\,
\frac{\kperpone^2 \kperptwo^2}{\kperpthree^2 \kperp^2}\, P_{a_1 a}\Big(\frac{x_1}{x_2+x_1}\Big)\,   
f_a(x_1+x_2,\kperpthreeb) \, ,
\end{equation}
where $f_a(x_1+x_2,\kperpthreeb)$ is the unintegrated single PDFs. Applying the method of References 
\cite{Kimber:1999xc,Kimber:2001sc}, this distribution
can be given the dependence on the hard scale $Q$,
\begin{equation}
\label{eq:13}
f_a(x_1+x_2,\kperpthree,Q) = T_a(Q,\kperpthree) \sum_{a'} \int_{x_1+x_2}^{1-\Delta} \frac{dz}{z} P_{aa'}(z,\kperpthree) \,
D_{a'}\!\left(\frac{x_1+x_2}{z},\kperpthree\right),
\end{equation}
where $\Delta=\kperpthree/Q$. Thus the  distribution (\ref{eq:12}) becomes scale dependent with equal scales, $Q_1=Q_2=Q$,
\begin{equation}
\label{eq:14}
f_{a_1 a_2}(x_1,x_2,\kperpone,\kperptwo,Q,Q) =  \, \frac{\alpha_s}{2\pi}  \frac{1}{x_1+x_2} \frac{\kperpone^2 \kperptwo^2}{\kperpthree^2 \kperp^2} 
P_{a_1 a}\!\left(\frac{x_1}{x_2+x_1}\right)   f_a(x_1+x_2,\kperpthree,Q) \; .
\end{equation}
The  reason for equal scales is that formula (\ref{eq:14}) only contains evolution of the unintegrated single parton density 
up to a scale $Q$ and then the splitting is treated with the  transverse momentum dependence.
The two partons from the splitting should evolve now. However, the initial partons have nonzero
transverse momenta which may be from the perturbative region, $\kperpone, \kperptwo \ge Q_0$. Thus, we should consider 
QCD radiation  with transverse momentum dependent splitting functions, see e.g. \cite{Hautmann:2012sh} for details. 
This stays, however, beyond the scope of the present analysis.

\section{CONCLUSIONS}

Following the method of Kimber, Martin and Ryskin (KMR) \cite{Kimber:1999xc,Kimber:2001sc}, 
we presented main points of the  construction of unintegrated double parton distribution functions which depend on parton transverse momenta, $\kperpone$ and 
$\kperptwo$, in addition to their two longitudinal momentum fractions, $x_1$ and $x_2$, and two factorization scales, $Q_1$ and $Q_2$.  We discussed  two contributions to the unintegrated DPDFs, corresponding to  the possibility that the two partons originate either from the proton or from the splitting of a single parton. In the first case the main formula is given by Equation~\ref{eq:8}. It corresponds to the fully perturbative domain
of transverse momenta, $\kperpone,\kperptwo \ge Q_0$. The formulae in the half-perturbative domains are presented in Reference~\cite{Golec-Biernat:2016vbt}.

In the perturbative case with parton splitting, we discussed two cases, the unfolding of the transverse momentum dependence from the last step in the DGLAP evolution of two partons, and the case where the transverse momenta are generated directly from  the single parton  splitting  into two partons. In the first case, only formula (\ref{eq:10}) with perturbative transverse
momenta makes sense. In the second case, we propose formula (\ref{eq:14}) 
which includes transverse momentum dependence generated from the perturbative splitting of one parton into two daughter partons. In that case, the KMR prescription is applied to the single PDF, in order to introduce the transverse momentum dependence, and then the splitting is treated by including the transverse momentum dependence. We kept the derivation in the strong ordering approximation to be consistent with the rest of the framework. The discussion of more subtle aspects of the transverse momentum dependence of UDPDFs, like
the dependence on an additional transverse momentum $\rperp$ or the transverse momentum dependence of the evolution after the parton splitting, as well as a numerical analysis,
have to be postponed to future publications.

\section{ACKNOWLEDGMENTS}
KGB would like to thank the organizers of "Diffraction 2016" for a fantastic atmosphere during the conference.
This work was supported by the National Science Center, Poland, Grant No. 2015/17/B/ST2/01838,   
by the Department of Energy  Grant No. DE-SC-0002145 and by the Center
for Innovation and Transfer of Natural Sciences and Engineering Knowledge in Rzesz\'ow.

\nocite{*}
\bibliographystyle{aipnum-cp}%
\bibliography{golec_diff16}

\begin{thebibliography}{13}%
\makeatletter
\providecommand \@ifxundefined [1]{%
 \@ifx{#1\undefined}
}%
\providecommand \@ifnum [1]{%
 \ifnum #1\expandafter \@firstoftwo
 \else \expandafter \@secondoftwo
 \fi
}%
\providecommand \@ifx [1]{%
 \ifx #1\expandafter \@firstoftwo
 \else \expandafter \@secondoftwo
 \fi
}%
\providecommand \natexlab [1]{#1}%
\providecommand \enquote  [1]{``#1''}%
\providecommand \bibnamefont  [1]{#1}%
\providecommand \bibfnamefont [1]{#1}%
\providecommand \citenamefont [1]{#1}%
\providecommand \href@noop [0]{\@secondoftwo}%
\providecommand \href [0]{\begingroup \@sanitize@url \@href}%
\providecommand \@href[1]{\@@startlink{#1}\@@href}%
\providecommand \@@href[1]{\endgroup#1\@@endlink}%
\providecommand \@sanitize@url [0]{\catcode `\$12\catcode `\&12\catcode
  `\#12\catcode `\^12\catcode `\_12\catcode `\%12\relax}%
\providecommand \@@startlink[1]{}%
\providecommand \@@endlink[0]{}%
\providecommand \url  [0]{\begingroup\@sanitize@url \@url }%
\providecommand \@url [1]{\endgroup\@href {#1}{\urlprefix }}%
\providecommand \urlprefix  [0]{URL }%
\providecommand \Eprint [0]{\href }%
\providecommand \doibase [0]{http://dx.doi.org/}%
\providecommand \selectlanguage [0]{\@gobble}%
\providecommand \bibinfo  [0]{\@secondoftwo}%
\providecommand \bibfield  [0]{\@secondoftwo}%
\providecommand \translation [1]{[#1]}%
\providecommand \BibitemOpen [0]{}%
\providecommand \bibitemStop [0]{}%
\providecommand \bibitemNoStop [0]{.\EOS\space}%
\providecommand \EOS [0]{\spacefactor3000\relax}%
\providecommand \BibitemShut  [1]{\csname bibitem#1\endcsname}%
\let\auto@bib@innerbib\@empty
\bibitem [{\citenamefont {Diehl}, \citenamefont {Ostermeier},\ and\
  \citenamefont {Schafer}(2012)}]{Diehl:2011yj}%
  \BibitemOpen
  \bibfield  {author} {\bibinfo {author} {\bibfnamefont {M.}~\bibnamefont
  {Diehl}}, \bibinfo {author} {\bibfnamefont {D.}~\bibnamefont {Ostermeier}}, \
  and\ \bibinfo {author} {\bibfnamefont {A.}~\bibnamefont {Schafer}},\ }\href
  {\doibase 10.1007/JHEP03(2012)089} {\bibfield  {journal} {\bibinfo  {journal}
  {JHEP}\ }\textbf {\bibinfo {volume} {1203}},\ p.\ \bibinfo {pages} {089}
  (\bibinfo {year} {2012})},\ \Eprint {http://arxiv.org/abs/1111.0910}
  {arXiv:1111.0910 [hep-ph]} \BibitemShut {NoStop}%
\bibitem [{\citenamefont {Abe}\ \emph {et~al.}(1997{\natexlab{a}})\citenamefont
  {Abe} \emph {et~al.}}]{Abe:1997bp}%
  \BibitemOpen
  \bibfield  {author} {\bibinfo {author} {\bibfnamefont {F.}~\bibnamefont
  {Abe}} \emph {et~al.} (\bibinfo {collaboration} {CDF Collaboration}),\ }\href
  {\doibase 10.1103/PhysRevLett.79.584} {\bibfield  {journal} {\bibinfo
  {journal} {Phys.Rev.Lett.}\ }\textbf {\bibinfo {volume} {79}},\ \unskip\
  \bibinfo {pages} {584--589} (\bibinfo {year}
  {1997}{\natexlab{a}})}\BibitemShut {NoStop}%
\bibitem [{\citenamefont {Abe}\ \emph {et~al.}(1997{\natexlab{b}})\citenamefont
  {Abe} \emph {et~al.}}]{Abe:1997xk}%
  \BibitemOpen
  \bibfield  {author} {\bibinfo {author} {\bibfnamefont {F.}~\bibnamefont
  {Abe}} \emph {et~al.} (\bibinfo {collaboration} {CDF Collaboration}),\ }\href
  {\doibase 10.1103/PhysRevD.56.3811} {\bibfield  {journal} {\bibinfo
  {journal} {Phys.Rev.}\ }\textbf {\bibinfo {volume} {D56}},\ \unskip\ \bibinfo
  {pages} {3811--3832} (\bibinfo {year} {1997}{\natexlab{b}})}\BibitemShut
  {NoStop}%
\bibitem [{\citenamefont {Abazov}\ \emph {et~al.}(2010)\citenamefont {Abazov}
  \emph {et~al.}}]{Abazov:2009gc}%
  \BibitemOpen
  \bibfield  {author} {\bibinfo {author} {\bibfnamefont {V.}~\bibnamefont
  {Abazov}} \emph {et~al.} (\bibinfo {collaboration} {D0 Collaboration}),\
  }\href {\doibase 10.1103/PhysRevD.81.052012} {\bibfield  {journal} {\bibinfo
  {journal} {Phys.Rev.}\ }\textbf {\bibinfo {volume} {D81}},\ p.\ \bibinfo
  {pages} {052012} (\bibinfo {year} {2010})},\ \Eprint
  {http://arxiv.org/abs/0912.5104} {arXiv:0912.5104 [hep-ex]} \BibitemShut
  {NoStop}%
\bibitem [{\citenamefont {Aad}\ \emph {et~al.}(2013)\citenamefont {Aad} \emph
  {et~al.}}]{Aad:2013bjm}%
  \BibitemOpen
  \bibfield  {author} {\bibinfo {author} {\bibfnamefont {G.}~\bibnamefont
  {Aad}} \emph {et~al.} (\bibinfo {collaboration} {ATLAS Collaboration}),\
  }\href {\doibase 10.1088/1367-2630/15/3/033038} {\bibfield  {journal}
  {\bibinfo  {journal} {New J.Phys.}\ }\textbf {\bibinfo {volume} {15}},\ p.\
  \bibinfo {pages} {033038} (\bibinfo {year} {2013})},\ \Eprint
  {http://arxiv.org/abs/1301.6872} {arXiv:1301.6872 [hep-ex]} \BibitemShut
  {NoStop}%
\bibitem [{\citenamefont {Chatrchyan}\ \emph {et~al.}(2014)\citenamefont
  {Chatrchyan} \emph {et~al.}}]{Chatrchyan:2013xxa}%
  \BibitemOpen
  \bibfield  {author} {\bibinfo {author} {\bibfnamefont {S.}~\bibnamefont
  {Chatrchyan}} \emph {et~al.} (\bibinfo {collaboration} {CMS Collaboration}),\
  }\href {\doibase 10.1007/JHEP03(2014)032} {\bibfield  {journal} {\bibinfo
  {journal} {JHEP}\ }\textbf {\bibinfo {volume} {1403}},\ p.\ \bibinfo {pages}
  {032} (\bibinfo {year} {2014})},\ \Eprint {http://arxiv.org/abs/1312.5729}
  {arXiv:1312.5729 [hep-ex]} \BibitemShut {NoStop}%
\bibitem [{\citenamefont {Aad}\ \emph {et~al.}(2014)\citenamefont {Aad} \emph
  {et~al.}}]{Aad:2014rua}%
  \BibitemOpen
  \bibfield  {author} {\bibinfo {author} {\bibfnamefont {G.}~\bibnamefont
  {Aad}} \emph {et~al.} (\bibinfo {collaboration} {ATLAS Collaboration}),\
  }\href {\doibase 10.1007/JHEP04(2014)172} {\bibfield  {journal} {\bibinfo
  {journal} {JHEP}\ }\textbf {\bibinfo {volume} {1404}},\ p.\ \bibinfo {pages}
  {172} (\bibinfo {year} {2014})},\ \Eprint {http://arxiv.org/abs/1401.2831}
  {arXiv:1401.2831 [hep-ex]} \BibitemShut {NoStop}%
\bibitem [{\citenamefont {Kimber}, \citenamefont {Martin},\ and\ \citenamefont
  {Ryskin}(2000)}]{Kimber:1999xc}%
  \BibitemOpen
  \bibfield  {author} {\bibinfo {author} {\bibfnamefont {M.~A.}\ \bibnamefont
  {Kimber}}, \bibinfo {author} {\bibfnamefont {A.~D.}\ \bibnamefont {Martin}},
  \ and\ \bibinfo {author} {\bibfnamefont {M.~G.}\ \bibnamefont {Ryskin}},\
  }\href {\doibase 10.1007/s100520000326} {\bibfield  {journal} {\bibinfo
  {journal} {Eur. Phys. J.}\ }\textbf {\bibinfo {volume} {C12}},\ \unskip\
  \bibinfo {pages} {655--661} (\bibinfo {year} {2000})},\ \Eprint
  {http://arxiv.org/abs/hep-ph/9911379} {arXiv:hep-ph/9911379 [hep-ph]}
  \BibitemShut {NoStop}%
\bibitem [{\citenamefont {Kimber}, \citenamefont {Martin},\ and\ \citenamefont
  {Ryskin}(2001)}]{Kimber:2001sc}%
  \BibitemOpen
  \bibfield  {author} {\bibinfo {author} {\bibfnamefont {M.~A.}\ \bibnamefont
  {Kimber}}, \bibinfo {author} {\bibfnamefont {A.~D.}\ \bibnamefont {Martin}},
  \ and\ \bibinfo {author} {\bibfnamefont {M.~G.}\ \bibnamefont {Ryskin}},\
  }\href {\doibase 10.1103/PhysRevD.63.114027} {\bibfield  {journal} {\bibinfo
  {journal} {Phys. Rev.}\ }\textbf {\bibinfo {volume} {D63}},\ p.\ \bibinfo
  {pages} {114027} (\bibinfo {year} {2001})},\ \Eprint
  {http://arxiv.org/abs/hep-ph/0101348} {arXiv:hep-ph/0101348 [hep-ph]}
  \BibitemShut {NoStop}%
\bibitem [{\citenamefont {Golec-Biernat}\ and\ \citenamefont
  {Stasto}(2016)}]{Golec-Biernat:2016vbt}%
  \BibitemOpen
  \bibfield  {author} {\bibinfo {author} {\bibfnamefont {K.}~\bibnamefont
  {Golec-Biernat}}\ and\ \bibinfo {author} {\bibfnamefont {A.~M.}\ \bibnamefont
  {Stasto}},\ }\href@noop {} {\  (\bibinfo {year} {2016})},\ \Eprint
  {http://arxiv.org/abs/1611.02033} {arXiv:1611.02033 [hep-ph]} \BibitemShut
  {NoStop}%
\bibitem [{\citenamefont {Snigirev}(2003)}]{Snigirev:2003cq}%
  \BibitemOpen
  \bibfield  {author} {\bibinfo {author} {\bibfnamefont {A.~M.}\ \bibnamefont
  {Snigirev}},\ }\href {\doibase 10.1103/PhysRevD.68.114012} {\bibfield
  {journal} {\bibinfo  {journal} {Phys. Rev.}\ }\textbf {\bibinfo {volume}
  {D68}},\ p.\ \bibinfo {pages} {114012} (\bibinfo {year} {2003})},\ \Eprint
  {http://arxiv.org/abs/hep-ph/0304172} {arXiv:hep-ph/0304172} \BibitemShut
  {NoStop}%
\bibitem [{\citenamefont {Ceccopieri}(2011)}]{Ceccopieri:2010kg}%
  \BibitemOpen
  \bibfield  {author} {\bibinfo {author} {\bibfnamefont {F.~A.}\ \bibnamefont
  {Ceccopieri}},\ }\href {\doibase 10.1016/j.physletb.2011.02.047} {\bibfield
  {journal} {\bibinfo  {journal} {Phys. Lett.}\ }\textbf {\bibinfo {volume}
  {B697}},\ \unskip\ \bibinfo {pages} {482--487} (\bibinfo {year} {2011})},\
  \Eprint {http://arxiv.org/abs/1011.6586} {arXiv:1011.6586 [hep-ph]}
  \BibitemShut {NoStop}%
\bibitem [{\citenamefont {Hautmann}, \citenamefont {Hentschinski},\ and\
  \citenamefont {Jung}(2012)}]{Hautmann:2012sh}%
  \BibitemOpen
  \bibfield  {author} {\bibinfo {author} {\bibfnamefont {F.}~\bibnamefont
  {Hautmann}}, \bibinfo {author} {\bibfnamefont {M.}~\bibnamefont
  {Hentschinski}}, \ and\ \bibinfo {author} {\bibfnamefont {H.}~\bibnamefont
  {Jung}},\ }\href {\doibase 10.1016/j.nuclphysb.2012.07.023} {\bibfield
  {journal} {\bibinfo  {journal} {Nucl. Phys.}\ }\textbf {\bibinfo {volume}
  {B865}},\ \unskip\ \bibinfo {pages} {54--66} (\bibinfo {year} {2012})},\
  \Eprint {http://arxiv.org/abs/1205.1759} {arXiv:1205.1759 [hep-ph]}
  \BibitemShut {NoStop}%
\end{thebibliography}%

\end{document}